\documentclass[pra,aps,twocolumn]{revtex4-2}
\usepackage{bm,dcolumn,amsmath,graphicx,amsfonts,amssymb,lipsum,mathtools}
\usepackage{epsfig}
\usepackage{tikz}
\usepackage{xcolor}

\usepackage{graphicx,subfigure}

\begin{document}

	\title{High-accuracy optical clocks with sensitivity to the fine-structure constant variation based on Sm$ ^{10+} $}

	\author{Saleh O. Allehabi$^{1}$, V. A. Dzuba$^2$, and V. V. Flambaum$^2$}
	
	\affiliation{$^{1}$Department of Physics, Faculty of Science, Islamic University of Madinah, Madinah 42351, Kingdom of Saudi Arabia}
	\affiliation{$^2$School of Physics, University of New South Wales, Sydney 2052, Australia}

	\date{\today}
	
	\begin{abstract}
		We identify two metastable excited states in Sm$ ^{10+} $ highly charged ion as candidates for high accuracy optical clocks. Several atomic properties relevant to optical clock development are calculated using relativistic many-body methods. This includes energy levels, transition amplitudes, lifetimes, scalar polarizabilities, black body radiation shift, and the sensitivity to the fine structure constant variation. We found that the clock transitions are not sensitive to perturbation, e.g., relative black body radiation shifts  are $\sim 10^{-19}$. The enhancement factor for the $\alpha$ variation is $\sim$ 0.8 for one clock transition and  $\sim$ 16 for another.
		
	\end{abstract}
	
	\maketitle

	\section{Introduction}
	
In recent times, there has been significant progress in the accuracy of frequency measurements in optical clock transitions of various atomic systems, including Sr, Yb, Al$ ^{+} $, Hg, Hg$ ^{+} $, and Yb$ ^{+} $. The fractional precision of these measurements has reached a remarkable level of  $10 ^{-18}$ ~\cite{ohmae2021transportable,brewer2019al+,huntemann2016single,mcgrew2018atomic,nicholson2015systematic,nemitz2016frequency,bloom2014optical,middelmann2012high,ludlow2018optical,PhysRevResearch.3.L042036}. 
This progress has led to increased interest in finding clock transitions which combine high accuracy of the measurements with high sensitivity to the manifestations of new physics, such as e.g. the variation of the fundamental constants and dark matter which may produce this variation.

	Highly charged ions (HCI) have been proposed as a way to search for optical transitions that are sensitive to fine structure constant variations over time~\cite{berengut2010enhanced}. HCI can be used to measure the relative change of the clock frequencies which then can be translated to the time variation of the fine structure constants 
 thereby revealing new physics~\cite{berengut2011electron,berengut2012optical,kozlov2018highly}.
The key to find optical transitions in HCI sensitive to variation of the fine structure constant is  by {\em level crossing} (when energies of the states are considered as functions of the ionization degree). Highest sensitivity can be found in ions with high $Z$ and ionization degree $Z_i$ in transitions which in single-electron approximation can be considered as  $s-f$ or $p-f$ transitions~\cite{berengut2010enhanced}.
The search for such transitions is now a popular area of research   (see e.g.  reviews~\cite{kozlov2018highly,Sahoo}).

	

\begin{figure}[tb]
		\centering
		
		{\includegraphics[width=0.42\textwidth]{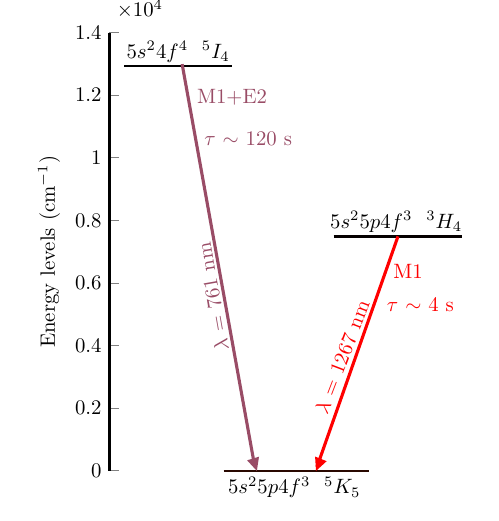}} 
		
\caption{The first three states of the Sm$^{10+}$ ion. Both excited states are metastable and corresponding transitions to the ground state are considered as clock transitions. Dominating contributions are indicated by symbols M1 (magnetic dipole) and E2 (electric quadrupole). Note that the M1 contribution to the transition at $\lambda =  761$~nm is strongly suppressed due to the difference in leading configurations. Therefore, the electric quadrupole also gives significant contribution.}
	\label{f:1}
\end{figure}

	
The Sm$ ^{+10} $ ion is examined in this work as a potential candidate for next-generation atomic clock. The ion has several advantages.
Two excited metastable states are considered as clock states (see Fig.~\ref{f:1}). Corresponding transitions between ground and metastable states have different sensitivity to the variation of the fine structure constant because one is the transition between states of the same configuration while the other correspond to the $p-f$ single-electron transition.
A change of fine-structure constant can be monitored by measuring one frequency versus another over long period of time (see, e.g. ~\cite{dzuba1999space,dzuba1999calculations,Science.319}). Samarium has seven stable isotopes, five of which have no nuclear spin, which makes it ideal for studying King plot nonlinearities. For a study of this kind, at least four stable isotopes and two transitions with potentially high accuracy of measurements are required. This may lead to useful  information about nuclear structure, and limit the strength of new interactions mediated by scalar bosons~\cite{berengut2018probing,flambaum2018isotope,delaunay2017probing}.
The transitions are not sensitive to external perturbations, e.g. the relative blackbody radiation shift (BBR) is small, $\sim 10^{-19}$ and can be further suppressed by combining two frequencies~\cite{yudin2011atomic}. 
	
The calculations of this work were performed using the configuration interaction with single-double coupled clusters (CI+SD)~\cite{dzuba2014combination} method and the configuration interaction with perturbation theory (CIPT)~\cite{dzuba2017combining} method. We calculated energy levels, Land\'{e} g-factors, transition amplitudes between low-lying states, quadrupole moments, and clock state lifetimes. For the estimation of the blackbody radiation shifts of clock frequencies, we also calculated the scalar polarizabilities of the ground and excited clock states. 
We also calculated the sensitivity of the clock transitions to the variation of the fine structure constant.

	\section{Method}
	
	\subsection{Calculation of energy levels}
	
The ground state configuration of Sm$ ^{10+} $ ion is [Pd]$5s^25p4f^3$. We consider it as a system with Pd-like closed-shell core with six valence electrons above it. Thus, we attribute the $5s$ electrons to the valence space. We use a combination of configuration interaction (CI) with linearized single-double coupled cluster (SD) methods~\cite{dzuba2014combination}. The calculations are performed using the $V^{N-M } $ approximation~\cite{dzuba2005v}, where $N$ is the total number of electrons and $M$ is the number of valence electrons ($N=46$, $M=6$ in our case).
The initial relativistic Hartree-Fock (RHF) procedure is done for the Pd-like closed-shell Sm$^{16+} $ ion.
The RHF Hamiltonian includes the Breit interaction and the quantum electrodynamic (QED) corrections,
	\begin{eqnarray} \label{e:RHF}
		&&\hat H^{\rm RHF}= c\bm{\alpha}\cdot\mathbf{p}+(\beta -1)mc^2+V_{\rm nuc}(r)+ \\
		&&V_{\rm core}(r) + V_{\rm Breit}(r) + V_{\rm QED}(r), \nonumber
	\end{eqnarray}
	where $c$ is the speed of light, $\bm{\alpha}$ and $\beta$ are the Dirac matrices, \textbf{p} is the electron momentum, $m$ is the electron mass, $V_{\rm nuc}$ is the nuclear potential obtained by integrating the Fermi distribution of the nuclear charge density,  $V_{\rm core}(r)$ is the self-consistent RHF potential created by the electrons of the closed-shell core, $V_{\rm Breit}(r)$ is Breit potential~\cite{DzuFlaSaf06},
	 $V_{\rm QED}(r)$ is a model QED potential~\cite{radpot}.

After completing the self-consistent procedure for the core, the B-spline technique~\cite{johnson1986computation,johnson1988finite} is used to create a complete set of single-electron wave functions. The functions are constructed as linear combinations of B-splines, which are eigenstates of the RHF Hamiltonian. We use 40 B-splines of order 9 in a box with a radius of $R_{\rm max}=40a_B$; the orbital angular momentum 0~$\leq$~\textit{l}~$\leq$~6.
These basis states are used to solve the SD equations for the core and for the valence states~\cite{dzuba2014combination} and for constructing the many-electron basis states for the CI calculations. 

Solving the SD equations gives us two correlation operators, $\Sigma_1$ and $\Sigma_2$. $\Sigma_1$ describes the correlation interaction between a particular valence electron and the core, whereas $\Sigma_2$ describes the Coulomb interaction screening between a pair of valence electrons~\cite{dzuba1987correlation,dzuba1996combination,dzuba2005v}.


The effective CI+SD Hamiltonian has the form
\begin{eqnarray} \label{e:HCI}
	\hat H^{\rm CI}=\sum_{i=1}^{M} \left(\hat H^{\rm RHF}+\Sigma_1\right)_i 
	+\sum_{i<j}^{M} \left(\dfrac{e^2}{|r_i-r_j|}+ \Sigma_{2ij}\right)
\end{eqnarray}
Here $i$ and $j$ enumerate valence electrons, summation goes over valence electrons, $ e $ is the electron charge, and $r$ is the position operator of the electrons. Note that Breit interaction is included in first term of (\ref{e:HCI}) ($\hat H^{\rm RHF}$) but not included in the second term. The contribution of the Breit interaction to the second term is small. It is smaller than the contribution  of higher order terms  in the calculation of  $\Sigma_{2ij}$.

There is a well-known fact that increasing the number of valence electrons exponentially increases the size of the CI matrix. We have six valence electrons, resulting in a matrix with an exceptionally large size. It will require considerable computational power to handle such matrix. In exchange for some accuracy, the size of the CI matrix can be reduced by orders of magnitude by using the CIPT (configuration interaction with perturbation theory) method suggested in~\cite{dzuba2017combining}. By dividing many-electron basis states into two large groups, low-energy states and high-energy states and ignoring the off-diagonal matrix elements between high-energy states the CI equations can be reduced to


\begin{equation}
	\langle i|H^\mathrm{eff}|j\rangle = \langle i|H^\mathrm{CI}|j\rangle+\sum_{k}\frac{\langle i|H^\mathrm{CI}|k\rangle\langle k|H^\mathrm{CI}|j\rangle}{E-E_{k}}.
	\label{e:CIPT}
\end{equation}

Here, $H^\mathrm{CI}$ is given by (\ref{e:HCI}), $i, j \leqslant N_{\text {eff }}, N_{\text {eff }}<k \leqslant N_{\text {total}}$, 
$N_{\text {eff }}$ is the number of low-energy states and $N_{\text {total}}$ is the total number of many-electron basis states.
Note that the choice  of $N_{\text {eff }}$ is arbitrary. One can increase it until the results are stable. In our calculations $N_{\text {eff }} \sim 10^4$ and $N_{\text {total}} \sim 10^7$.

Parameter $E$ in (\ref{e:CIPT}) is the energy of the state of interest, and $E_k$ denotes the diagonal matrix element for high-energy states, $E_{k}=\left\langle k\left|H^{\mathrm{CI}}\right| k\right\rangle$. The summation in (\ref{e:CIPT}) runs over all high-energy states. 
Note that the parameter $E$ in the denominator of (\ref{e:CIPT}) is the same as the energy of the state of interest which is to be obtained from solving the CI equations. Since this energy is not known in advance, iterations over $E$ are needed to find it.
More detailed explanations of the technique can be found in Ref.~\cite{dzuba2017combining}.


Knowing the values of the state's $g$-factors together with their energies, total angular momentum ($J$) and parity is very useful for the states identification.  Therefore, we calculate the $g$-factors as an expectation value of the magnetic dipole (M1) operator.
Comparing the result with the non-relativistic expression
\begin{equation}\label{e:g}
	g_{NR} = 1 + \frac{J(J+1)-L(L+1)+S(S+1)}{2J(J+1)}
\end{equation}
helps to label the states with standard notations $^{(2S+1)}L_J$. Here the values of total spin ($S$) and orbital momentum ($L$) are chosen to fit the calculated values of the $g$-factors. Not that this procedure is not unambiguous since the $SLJ$ scheme does not work well for heavy ions like Sm$^{10+}$. 
We just use it for labelling of the states in the absence of obvious better alternative.

\subsection{Calculation of transition probabilities and lifetimes}

To determine the values of the transition probabilities (T) of the atomic states, we need to calculate the transition amplitudes ($A$). To do this, we use the time-dependent Hartree-Fock (TDHF) method, which is equivalent to the random-phase approximation (RPA). A TDHF equation for the core can be written as follows:
\begin{equation}\label{e:RPA}
	\left(\hat H^{\rm RHF}-\epsilon_c\right)\delta\psi_c=-\left(\hat f+\delta V^{f}_{\rm core}\right)\psi_c.
\end{equation}
Here, $ \hat H^{\rm RHF} $ is the RHF Hamiltonian (\ref{e:RHF}), index $c$ numerates states in the core, $\epsilon_c$ is the energy of electron state $ c $, and $\psi_c$ is the  wave function of the state, $\delta\psi_c$ is a correction to the wave function due to the external field $\hat f$, 
$\delta V^{f}_{\rm core}$ is the correction to the self-consistent RHF potential of the core 
generated by the modifications to all core states in the external field. 

After solving the RPA equations  (\ref{e:RPA}) self-consistently for finding $\delta V^{f}_{\rm core}$, the off-diagonal matrix element then gives the transition amplitude between valence states 
\begin{equation}\label{e:Ab}
	A_{i j}=\left\langle\Psi_i\left|\hat{f}+\delta V^{f}_{\rm core }\right| \Psi_j\right\rangle
\end{equation}
Here, $\left|\Psi_i\right\rangle,\left|\Psi_j\right\rangle$ are the many-electron wave functions obtained by solving Eq.~(\ref{e:CIPT}).


Once the transition amplitudes have been obtained, the transition probabilities can be determined using the following formulae (atomic units are assumed):
\begin{equation}\label{e:Td}
	T^{M1}_{ab} = \frac{4}{3}(\alpha\omega_{ab})^3 \frac{A^2_{M1}}{2J_b+1},
\end{equation}

\begin{equation}\label{e:Tq2}
	T^{\rm E2}_{ab} = \frac{1}{15}(\alpha\omega_{ab})^5 \frac{A^2_{E2}}{2J_b+1},
\end{equation}


We consider two types of transitions: 
magnetic dipole ($T_{M1}$) transition (\ref{e:Td}), and electric quadrupole ($T_{E2}$) transition (\ref{e:Tq2}). In both equations, $\alpha$ is the fine structure constant ($\alpha\approx\frac{1}{137}$), $\omega_{ab}$ is the energy difference between the lower ($ a $) and upper ($ b $) states, \textit{A} is the transition amplitude (reduced matrix element)  derived from Eq. (\ref{e:Ab}), and $J_b$ is the total angular momentum of the upper state \textit{b}. Note that magnetic amplitudes $A_{M1}$ contain the Bohr magneton $\mu_B$ ($\mu_B = \alpha/2 \approx 3.65 \times 10^{-3}$ in the Gauss-type atomic units). 

The lifetime (in seconds) of the upper state $b$ is given by
\begin{equation}\label{e:tau}
	\tau_b =  2.4189 \times 10^{-17}/\sum_a T_{ab},
\end{equation}
where the summation goes over all possible transitions to lower states $a$.

\subsection{Accuracy of the calculations}

The accuracy of the present approach was discussed in details in our earlier works (see, e.g. \cite{allehabi2022atomic,allehabi2022high,allehabi2023prospects}).
The most relevant discussion can be found in Ref.~\cite{allehabi2022high} where a number of highly charged ions have been considered.
It was demonstrated  there that the accuracy of the results ranges from 1\% to 10\%. Given that in present work we consider similar atomic systems, the accuracy in the present work is also similar.

\section{RESULTS AND DISCUSSIONS}

\subsection{Energy levels, g-factors, transition amplitudes, and lifetimes}

We present the energy levels, $g$-factors and lifetimes of some states  of the Sm$^{10+}$ ion in Table~\ref{t:Energy}. 
First two long-living excited states are considered as clock states.
The energies and $g$-factors are presented for 29 lowest even states with  $1 \leq J \leq 8$.

The odd states have also been calculated, and it was found that the lowest odd excited state, that connects to the ground state via an electric dipole (E1) transition, is the 5s5p4f$ ^{4} $ $ ^{5}K_{4} $ state with the energy of $E = 158874~\rm{cm}^{-1}$ which is well outside of the optical region. 
	There are no good cooling transitions in the ion,  and a sympathetic cooling method~\cite{schmidt2005spectroscopy,ludlow2015optical} can be used instead. This method involves co-trapping the target ions with auxiliary ions, known as logic ions, which can be efficiently cooled using laser techniques. The energy transferred from the logic ions to the target ions cools the entire system. For optimal sympathetic cooling, it is essential that the charge (q)-to-mass (m) ratio of the clock ion is similar to that of the logic ion~\cite{wubbena2012sympathetic}. This similarity ensures efficient energy transfer and minimizes differential motion between the two species, leading to more effective overall cooling.
For the ion considered, the (q/m)$ _{clock~ion} $ ratio is 0.0665. Given this, the single Mg$ ^{+} $ ion, with a (q/m)$ _{logic~ion} $ ratio of 0.0411, is estimated to be the most suitable logic ion.


As shown in Table~\ref{t:Tran}, we have calculated the transition amplitudes (A) and transition probabilities (T) for six low states of the Sm$^{10+}$ ion.
Since all low states have the same parity, only magnetic dipole (M1) and electric quadrupole (E2) contributions are significant.
By using Eq.~(\ref{e:tau}), we derived the lifetimes ($\tau$) of the excited states using the transition rates from Table~\ref{t:Tran}. 
\begin{table}[]
	\caption{\label{t:Energy}
		Excitation energies ($E$, cm$^{-1}$), $g$-factors,  and lifetimes for low-lying states of the Sm$ ^{10+} $ ion.} 
	\begin{ruledtabular}
		\begin{tabular}{cccccc}
			
			\multicolumn{1}{c}{ No. }&
			
			\multicolumn{1}{c}{ Conf. }&
			\multicolumn{1}{c}{  LSJ   }&
			\multicolumn{1}{c}{ Energy  }&
			\multicolumn{1}{c}{ $g$ }&
			\multicolumn{1}{c}{ $\tau$   }\\
			
			\hline
			
1&    5s$ ^{2} $5p4f$ ^{3} $&         $ ^{5} K    _{5} $&        0.00&   0.7469   &\\
			
2&      5s$ ^{2} $5p4f$ ^{3} $&        $  ^{3}H_{4} $  &   7883 &  0.7410   &4.03 s\\
3&   5s$ ^{2} $4f$ ^{4} $&             $  ^{5}I_{4} $    & 13139 & 0.6670&120.1 s    \\
4&   5s$ ^{2} $5p4f$ ^{3} $&           $  ^{5}K_{6} $ &  13291  & 0.9426   &10.03 ms\\
5&   5s$ ^{2} $5p4f$ ^{3} $&           $  ^{5}H_{5} $ &  16880  & 0.9711   &31.84 ms\\
6&  5s$ ^{2} $5p4f$ ^{3} $&         $  ^{3}I_{7} $  & 17214  & 1.1589   &0.48 s\\
7&  5s$ ^{2} $5p4f$ ^{3} $&           $  ^{5}G_{2} $   & 18525 & 0.6117&\\
8&  5s$ ^{2} $5p4f$ ^{3} $&           $  ^{5}K_{8} $   & 19447 & 1.1825&\\
9&  5s$ ^{2} $5p4f$ ^{3} $&         $  ^{5}K_{7} $  & 20337 &  1.1519   &\\
10&   5s$ ^{2} $4f$ ^{4} $&             $  ^{5}I_{8} $    & 20831&  1.2413&    \\
11&   5s$ ^{2} $5p4f$ ^{3} $&           $  ^{3}H_{6} $ &  20980&   1.1312  &\\
		
12&   5s$ ^{2} $4f$ ^{4} $&             $  ^{5}I_{5} $    & 23156  & 0.9092 &    \\
		
13&   5s$ ^{2} $5p4f$ ^{3} $&           $  ^{3}D_{1} $ &  23204  & 0.6597&\\

14&   5s$ ^{2} $5p4f$ ^{3} $&           $  ^{3}G_{3} $ &  24770  & 0.7638&\\

15&   5s$ ^{2} $4f$ ^{4} $&             $  ^{5}I_{6} $    & 26355 &  1.0683 &    \\

16&   5s$ ^{2} $5p4f$ ^{3} $&           $  ^{1}H_{5} $ &  26839 &  0.9644&\\

17&   5s$ ^{2} $4f$ ^{4} $&             $  ^{5}I_{7} $    & 27057 &  1.1721 &    \\

18&   5s$ ^{2} $5p4f$ ^{3} $&           $  ^{5}F_{2} $ &  27529 &  0.8476&\\

19&   5s$ ^{2} $5p4f$ ^{3} $&           $  ^{5}G_{3} $ &  28248 &  0.8296&\\

20&   5s$ ^{2} $5p4f$ ^{3} $&           $  ^{5}G_{4} $ &  28260 &  1.0888&\\

21&   5s$ ^{2} $5p4f$ ^{3} $&           $  ^{3}S_{1} $ &  29074 &  2.0820 &\\

22&   5s$ ^{2} $5p4f$ ^{3} $&           $  ^{5}F_{2} $ &  29458  & 1.0945 &\\

23&   5s$ ^{2} $5p4f$ ^{3} $&           $  ^{3}G_{4} $ &  31509 &  1.0703&\\

24&   5s$ ^{2} $5p4f$ ^{3} $&           $  ^{1}D_{2} $ &   32264  & 0.9642 &\\

25&   5s$ ^{2} $5p4f$ ^{3} $&           $  ^{3}F_{4} $ &  32324  & 1.1831 &\\

26&   5s$ ^{2} $5p4f$ ^{3} $&           $  ^{5}F_{3} $ &  32520  & 1.2525 &\\

27&   5s$ ^{2} $5p4f$ ^{3} $&           $  ^{5}H_{5} $ &  33467  & 1.1459&\\

28&   5s$ ^{2} $5p4f$ ^{3} $&           $  ^{3}D_{2} $ &   33576  & 1.1598 &\\

29&   5s$ ^{2} $4f$ ^{4} $&             $  ^{5}F_{1} $    & 33695  & 0.1278&    \\

		\end{tabular}

	\end{ruledtabular}
\end{table}

\begin{table*}
	
	\caption{\label{t:Tran} Transition amplitudes (\textit{A}, a.u.) and transition probabilities (T, 1/s) for some low  states.} 
	
	\begin{ruledtabular}
		\begin{tabular}{cc c cc cc }
			&&&
			\multicolumn{2}{c}{($\omega$) }&&\\

			\cline{4-5}

			\multicolumn{1}{c}{Transition}& 
	 	\multicolumn{1}{c}{$J_b$}&
			\multicolumn{1}{c}{Type}&
			\multicolumn{1}{c}{ [cm$^{-1}$]}&
			\multicolumn{1}{c}{ [a.u.]}&

			\multicolumn{1}{c}{\textit{A} [a.u]}&
			\multicolumn{1}{c}{T [s$^{-1}$]}\\

			\hline

			2-1	&4&	M1&	7883&	0.0359&	$-1.50\times10^{-3}$&	0.2484\\
			2-1	&4	&E2&	7883	&0.0359&	$-0.2125$&	$1.708\times10^{-5}$\\	
			\\
			3-1	&4&	M1&	13139&	0.0599&	$1.14\times10^{-4}$&	$6.637\times10^{-3}$\\	
			3-1&	4&	E2&	13139&	0.0599&	-0.49522&	$1.194\times10^{-3}$\\	
			3-2	&4&	M1	&5257	&0.0240	&$1.19\times10^{-4}$&	$4.606\times10^{-4}$\\	
			3-2	&4&	E2&	5257&	0.0240&	-0.85906&	$3.682\times10^{-5}$\\	
			\\
			4-1	&6	&M1	&13291&	0.0606	&0.0163&	96.750\\
			4-1	&6	&E2&	13291&	0.0606&	0.35168&	$4.414\times10^{-4}$\\	
			4-2	&6&	E2&	5408&	0.0246&	-0.43874&	$7.664\times10^{-6}$	\\
			4-3	&6&	E2	&152&	0.0007&	-0.1065&	$7.801\times10^{-15}$\\	
			\\
			
			5-1	&5&	M1&	16880	&0.0769	&$2.22\times10^{-3}$&	4.352	\\
			5-1	&5	&E2&	16880&	0.0769&	$ -0.0491 $&	$3.361\times10^{-5}$\\	
			5-2	&5&	M1	&8998	&0.0410&	0.0142&	27.05	\\
			5-2	&5&	E2&	8998	&0.0410&	-0.0238&	$3.398\times10^{-7}$\\	
			5-3&	5	&M1	&3741	&0.0170	&$-8.92\times10^{-4}$	&$7.664\times10^{-3}$	\\
			5-3	&5&	E2&	3741&	0.0170&	0.22821&	$3.883\times10^{-7}$\\	
			5-4	&5&	M1	&3590	&0.0164&	$-7.58\times10^{-3}$&	0.4895	\\
			5-4	&5&	E2&	3590&	0.0164&	$-0.0697$&	$2.948\times10^{-8}$\\	
			\\
			
			6-1	&7	&E2	&17214	&0.0784	&$4.16\times10^{-3}$	&$1.953\times10^{-7}$\\
			6-4	&7&	M1&	3924&	0.0179&	$-0.0160$&	2.091\\	
			6-4	&7	&E2	&3924	&0.0179	&-0.21774&	$3.288\times10^{-7}$	\\
			6-5	&7&	E2&	334&	0.0015&	$-0.301$&	$2.799\times10^{-12}$\\

		\end{tabular}
	\end{ruledtabular}
\end{table*}

\begin{table*}
	\caption{\label{t:pol}
		Scalar static polarizabilities of the ground state, $\alpha_0({\rm GS})$, and clock states, $\alpha_0({\rm CS})$,  and BBR frequency shifts for the clock transitions.  $\delta\nu_{BBR}$/$\nu$ is the fractional  BBR shift; $\nu$ is the clock transition frequency. The values presented for $\alpha_0({\rm GS})$ and $\alpha_0({\rm CS})$ include core and valence contributions.	The value of the core polarizability is 0.246 $a_B^3$.
	}
	\begin{ruledtabular}
		\begin{tabular}{cc c cr lr}
			&&&

			\multicolumn{1}{c}{$\Delta \alpha (0)$} &
			\multicolumn{3}{c}{BBR, (\textit{T}= 300 K)} \\

			\cline{4-4}
			\cline{5-7}
			
			\multicolumn{1}{c}{Transition}&
			
			\multicolumn{1}{c}{$\alpha_0({\rm GS})$[$a_B^3$]}&
			\multicolumn{1}{c}{$\alpha_0({\rm CS})$[$a_B^3$]}&

			\multicolumn{1}{c}{$\alpha_0({\rm CS}) - \alpha_0({\rm GS})$} &
			
			\multicolumn{1}{c}{$\delta\nu_{BBR}$[Hz]}&
			\multicolumn{1}{c}{$\nu$[Hz]}&
			\multicolumn{1}{c}{$\delta\nu_{BBR}$/$\nu$} \\
			\hline
			
			
			2 $-$ 1  &1.4603&1.4544&$ - $0.0059&0.5081$\times 10^{-4}$&2.3631$\times 10^{14}$&$2.150$$\times 10^{-19}$ \\
			
			3 $-$ 1  &1.4603&1.7222&0.2619&$ - $0.2255$\times 10^{-2}$&3.9390$\times 10^{14}$&$-$5.725$\times 10^{-18}$ \\


		\end{tabular}
		
	\end{ruledtabular}
\end{table*}

\subsection{Polarizabilities and blackbody radiation shifts}

An important source of uncertainty in clock frequencies is blackbody radiation shift (BBR).
To estimate this shift one needs to know static scalar polarizability ($\alpha_v(0)$) for both states of the clock transition.

The polarizability of state $v$  can be expressed as a sum over a complete set of states connected to state $v$ via the electric dipole (E1) transitions 
\begin{equation}\label{e:pol}
	\alpha_v(0)=\dfrac{2}{3(2J_v+1)}\sum_{n}\frac{A_{vn}^2}{\omega_{vn}},
\end{equation}
where $J_v$ is the total angular momentum of state $v$ and $\omega_{vn}$ is the frequency of the transition. Notations $v$ and $n$ refer to many-electron atomic states. 

It is convenient to present the polarizability of an atomic system as a sum of the polarizability of the closed-shell core and the contribution of the valence electrons. For the closed-shell core, Eq.~(\ref{e:pol}) becomes
\begin{equation}\label{e:polCor}
	\alpha_v(0)=\frac{2}{3} \sum_c\left\langle v|\hat{d}| \delta \psi_c\right\rangle
\end{equation}
where $\hat{d}$ is the electric dipole moment operator, and $\delta \psi_c$ is the RPA correction to the core state $c$ [see Eq. (\ref{e:RPA})]. The summation goes over all states in the core. Then, Eq.~(\ref{e:pol}) is used for the contribution of the valence states.
We found that the value of the core polarizability of the Sm$^{10+}$ ion is  0.246 $a_B^3$.

For the calculations of the contributions of the valence electrons to the polarizabilities, we apply the technique developed in Ref.~\cite{dzuba2020calculation} for atoms or ions with open shells. The method relies on Eq.~(\ref{e:pol}) and the Dalgarno$ - $Lewis approach~\cite{dalgarno1955exact}, which reduces the summation in Eq.~(\ref{e:pol}) to
solving a matrix equation ( for more details, see Ref.~\cite{dzuba2020calculation}).

A comparison of the polarizabilities of the ground and excited clock states is presented in Table~\ref{t:pol}. The ground and the first excited clock state seem to have similar polarizabilities. The reason for this is that the frequency of the transition is small compared to the frequencies of the transitions to odd states ($E_{\rm odd} > 158000~\rm{cm}^{-1}$).
We have seen similar facts before in some other ions,
see Refs.~\cite{allehabi2023prospects,allehabi2022atomic,allehabi2022high}.


In atomic clocks, the BBR shift can considerably affect the frequency of the clock's transition. 
Its value in Hz is given by
\begin{equation}
	\delta \nu_{\rm BBR} = -1.063 \times  10^{-12}
		\left(\frac{\mathrm{T}(\mathrm{K})}{300}\right)^4\Delta \alpha(0),
\end{equation}

where $T$ is the temperature (room temperature is $ T $= 300 $ K $) and $ \Delta \alpha(0)= \alpha_0({\rm CS}) - \alpha_0({\rm GS})$ is the difference between the excited and ground state polarizabilities given in atomic units. 
A summary of the BBR shifts for the clock states investigated in this work is presented in Table~\ref{t:pol}. As can be seen, the relative BBR shifts for these transitions are among the smallest that have been examined so far,  $\sim 10^{-19}$. 

\subsection{Quadrupole moments}

The ground and clock states of the ion being considered have high values of total angular momentum (J= 5 and 4). This makes them susceptible to the interaction between the gradient of the external electric field and the quadrupole moments of the states, which causes a frequency shift. Hence, it is crucial to have knowledge about the quadrupole moment values.


The quadrupole interaction Hamiltonian  can be written as (see, e.g., Ref.~\cite{itano2000external})
\begin{equation}
	H_{\mathcal{Q}}=\sum_{q=-1}^1(-1)^q \nabla \mathcal{E}_q^{(2)} \hat{\Theta}_{-q},
\end{equation}
where $q$ is the operator component, and  the $\nabla \mathcal{E}_q^{(2)}$ tensor represents the gradient of the external electric field at the position of the system. $\hat{\Theta}_q$ is the electric quadrupole operator (the same as for the E2 transitions) and  describes as $\hat{\Theta}_q=r^2 C_q^{(2)}$, where $C_q^{(2)}$ is the normalized spherical function.

The electric-quadrupole moment $\Theta$ is defined as an expectation value of the $\hat{\Theta}_0$ operator,
\begin{equation}
	\begin{aligned}
		\Theta & =\left\langle n J J\left|\hat{\Theta}_0\right| n J J\right\rangle \\
		& =\langle n J\|\hat{\Theta}\| n J\rangle \sqrt{\frac{J(2 J-1)}{(2 J+3)(2 J+1)(J+1)}},
	\end{aligned}
\end{equation}
where $\langle n J\|\hat{\Theta}\| n J\rangle$ denotes the reduced matrix element (ME) of the electric-quadrupole operator. In Table~\ref{t:Q}, the reduced ME of the electric quadrupole operators for the considered states are shown along with their $\Theta$.

\begin{table}[!]
	\caption{\label{t:Q}
		Quadrupole moment ($\Theta$, a.u.) of the considered optical clock states.} 
	
	\begin{ruledtabular}
		\begin{tabular}{crcccc}
			
			\multicolumn{1}{c}{No.}& 
			\multicolumn{1}{c}{Conf.}&
			\multicolumn{1}{c}{LSJ}&
			
			\multicolumn{1}{c}{$E$ (cm$^{-1}$)}&
			
			\multicolumn{1}{c}{ ME (a.u.) }&
			\multicolumn{1}{c}{ $\Theta$ }\\&&&&

			\multicolumn{1}{c}{ $\left\langle J\|\Theta_{0}\| J\right\rangle$}&\\
			\hline
			
			1&    5s$ ^{2} $5p4f$ ^{3} $&         $ ^{5} K    _{5} $&  0000&     -0.3580&0.0820     \\
			2&      5s$ ^{2} $5p4f$ ^{3} $&        $  ^{3}H_{4} $  &   7882.62 &  -0.0611&-0.0145\\
			3&   5s$ ^{2} $4f$ ^{4} $&             $  ^{5}I_{4} $    &13139.18& 0.3950& 0.0940   \\
			
			
			

		\end{tabular}
		
	\end{ruledtabular}
	
\end{table}

\subsection{Sensitivity of the clock transitions to variation of the fine structure constant}


The sensitivity of atomic transitions to $\alpha$ variation is calculated by writing the frequencies of the transitions in the following form

\begin{equation}
	\omega(x)=\omega_0+q x
\end{equation}
Here $\omega_0$ is the present laboratory value of the transition frequency and $x=\left(\alpha / \alpha_0\right)^2-1$, where $\alpha_0$ is the physical value of $\alpha$, $q$ is the sensitivity coefficient, which can be determined by varying the value of $\alpha$ in computer codes and calculating numerical derivatives
\begin{equation}
	q \approx \frac{\omega(+\delta)-\omega(-\delta)}{2 \delta}
\end{equation}
To achieve linear behaviour, $\delta$ should be sufficiently small. On the other hand, it should be large enough to suppress numerical noise. In our case, we used  $\delta$ = 0.01 which leads to stable results. 

To study possible variation of the fine structure constant one needs to measure frequencies of at least two atomic transitions over a long period of time.
The ratio of changes in frequencies can be written as
\begin{equation} \label{e:w1w2}
	\delta\left(\frac{\omega_1}{\omega_2}\right)/\frac{\omega_1}{\omega_2}=
	\frac{\delta\omega_1}{\omega_1} - \frac{\delta\omega_2}{\omega_2} \equiv \left(K_1 - K_2 \right)\frac{\delta\alpha}{\alpha}.
\end{equation}

The $ K  $ parameter is known as the enhancement factor and has the formula $K=2q/\omega$ . 
It is clear that the highest sensitivity can be achieved if two atomic transitions have different sensitivity to the $\alpha$ variation.
For both studied clock transitions, $ q $ and $ K $ have been calculated and are summarized in Table~\ref{t:q}. 
As one can see, the sensitivities of the transitions to $\alpha$ variation are different. For these two transitions Eq.~(\ref{e:w1w2}) becomes
\begin{equation} \label{e:w11w22}
	\delta\left(\frac{\omega_1}{\omega_2}\right)/ \frac{\omega_1}{\omega_2}\approx 16 \frac{\delta\alpha}{\alpha}.
\end{equation}
For comparison, the highest sensitivity to $\alpha$ variation in working clocks can be found in the monitoring of the frequencies of E2 and E3 transitions in Yb$^+$, where $\Delta K \approx 7$ \cite{FlaDzuCJP}.
It  may be  convenient  that the measurements for two frequencies can be done with the use of the same ion.

\begin{table}[!]
	\caption{\label{t:q}
		Sensitivity of the clock transitions to the variation of the fine-structure constant ($q, K$) of Sm$^{10+}$.} 
	\begin{ruledtabular}
		\begin{tabular}{cccccc}

			\multicolumn{1}{c}{No.}&
			\multicolumn{1}{c}{Conf.}&
			\multicolumn{1}{c}{LSJ}&

			\multicolumn{1}{c}{$\omega$ (cm$^{-1}$)}&
			
			\multicolumn{1}{c}{$q$(cm$^{-1}$) }&
			\multicolumn{1}{c}{$K$}\\
			\hline

			2&      5s$ ^{2} $5p4f$ ^{3} $&        $  ^{3}H_{4} $  &   7882.62 & 3100 &0.79\\
			3&   5s$ ^{2} $4f$ ^{4} $&             $  ^{5}I_{4} $    &13139.18&108700  & 16.5   \\
			
			
			

		\end{tabular}
	\end{ruledtabular}
\end{table}

\section{Summary}

In this work, we demonstrated  that the Sm$ ^{10+} $ ion has two metastable states which can be used to build atomic optical clocks with low sensitivity to perturbations and high sensitivity to the hypothetical time variation of the fine structure constant.  Several atomic properties relevant to optical clock development have been calculated. These include energy levels, $g$-factors, lifetimes, electric quadrupole moments, BBR shift and sensitivity of clock frequencies to time variation of the fine structure constant. The relative value of the BBR shift is very small, $ \sim 10^{-19}$.
We found that if fine structure constant $\alpha$ changes with time, the ratio of the two clock frequencies changes 16 times faster.
This is significantly better than in best (in terms of sensitivity to $\alpha$ variation) operating atomic clocks based on the E3 and E2 clock transitions in the Yb$^+$ ion.


\begin{acknowledgments}

We are grateful to Alexander Kramida for the valuable discussion. This work was supported by the Australian Research Council Grants No. DP230101058 and DP200100150. SOA wishes to extend his sincere gratitude to the Deanship of Scientific Research at the Islamic University of Madinah (Kingdom of Saudi Arabia) for the support provided via the Post-Publishing Program.
	
\end{acknowledgments}

\bibliography{references3}

\begin{thebibliography}{41}
\expandafter\ifx\csname natexlab\endcsname\relax\def\natexlab#1{#1}\fi
\expandafter\ifx\csname bibnamefont\endcsname\relax
  \def\bibnamefont#1{#1}\fi
\expandafter\ifx\csname bibfnamefont\endcsname\relax
  \def\bibfnamefont#1{#1}\fi
\expandafter\ifx\csname citenamefont\endcsname\relax
  \def\citenamefont#1{#1}\fi
\expandafter\ifx\csname url\endcsname\relax
  \def\url#1{\texttt{#1}}\fi
\expandafter\ifx\csname urlprefix\endcsname\relax\def\urlprefix{URL }\fi
\providecommand{\bibinfo}[2]{#2}
\providecommand{\eprint}[2][]{\url{#2}}

\bibitem[{\citenamefont{Ohmae et~al.}(2021)\citenamefont{Ohmae, Takamoto,
  Takahashi, Kokubun, Araki, Hinton, Ushijima, Muramatsu, Furumiya, Sakai
  et~al.}}]{ohmae2021transportable}
\bibinfo{author}{\bibfnamefont{N.}~\bibnamefont{Ohmae}},
  \bibinfo{author}{\bibfnamefont{M.}~\bibnamefont{Takamoto}},
  \bibinfo{author}{\bibfnamefont{Y.}~\bibnamefont{Takahashi}},
  \bibinfo{author}{\bibfnamefont{M.}~\bibnamefont{Kokubun}},
  \bibinfo{author}{\bibfnamefont{K.}~\bibnamefont{Araki}},
  \bibinfo{author}{\bibfnamefont{A.}~\bibnamefont{Hinton}},
  \bibinfo{author}{\bibfnamefont{I.}~\bibnamefont{Ushijima}},
  \bibinfo{author}{\bibfnamefont{T.}~\bibnamefont{Muramatsu}},
  \bibinfo{author}{\bibfnamefont{T.}~\bibnamefont{Furumiya}},
  \bibinfo{author}{\bibfnamefont{Y.}~\bibnamefont{Sakai}},
  \bibnamefont{et~al.}, \bibinfo{journal}{Adv. Quantum Tech.}
  \textbf{\bibinfo{volume}{4}}, \bibinfo{pages}{2100015}
  (\bibinfo{year}{2021}).

\bibitem[{\citenamefont{Brewer et~al.}(2019)\citenamefont{Brewer, Chen, Hankin,
  Clements, Chou, Wineland, Hume, and Leibrandt}}]{brewer2019al+}
\bibinfo{author}{\bibfnamefont{S.~M.} \bibnamefont{Brewer}},
  \bibinfo{author}{\bibfnamefont{J.-S.} \bibnamefont{Chen}},
  \bibinfo{author}{\bibfnamefont{A.~M.} \bibnamefont{Hankin}},
  \bibinfo{author}{\bibfnamefont{E.~R.} \bibnamefont{Clements}},
  \bibinfo{author}{\bibfnamefont{C.-w.} \bibnamefont{Chou}},
  \bibinfo{author}{\bibfnamefont{D.~J.} \bibnamefont{Wineland}},
  \bibinfo{author}{\bibfnamefont{D.~B.} \bibnamefont{Hume}}, \bibnamefont{and}
  \bibinfo{author}{\bibfnamefont{D.~R.} \bibnamefont{Leibrandt}},
  \bibinfo{journal}{Phys. Rev. Lett.} \textbf{\bibinfo{volume}{123}},
  \bibinfo{pages}{033201} (\bibinfo{year}{2019}).

\bibitem[{\citenamefont{Huntemann et~al.}(2016)\citenamefont{Huntemann, Sanner,
  Lipphardt, Tamm, and Peik}}]{huntemann2016single}
\bibinfo{author}{\bibfnamefont{N.}~\bibnamefont{Huntemann}},
  \bibinfo{author}{\bibfnamefont{C.}~\bibnamefont{Sanner}},
  \bibinfo{author}{\bibfnamefont{B.}~\bibnamefont{Lipphardt}},
  \bibinfo{author}{\bibfnamefont{C.}~\bibnamefont{Tamm}}, \bibnamefont{and}
  \bibinfo{author}{\bibfnamefont{E.}~\bibnamefont{Peik}},
  \bibinfo{journal}{Phys. Rev. Lett.} \textbf{\bibinfo{volume}{116}},
  \bibinfo{pages}{063001} (\bibinfo{year}{2016}).

\bibitem[{\citenamefont{McGrew et~al.}(2018)\citenamefont{McGrew, Zhang,
  Fasano, Sch{\"a}ffer, Beloy, Nicolodi, Brown, Hinkley, Milani, Schioppo
  et~al.}}]{mcgrew2018atomic}
\bibinfo{author}{\bibfnamefont{W.}~\bibnamefont{McGrew}},
  \bibinfo{author}{\bibfnamefont{X.}~\bibnamefont{Zhang}},
  \bibinfo{author}{\bibfnamefont{R.}~\bibnamefont{Fasano}},
  \bibinfo{author}{\bibfnamefont{S.}~\bibnamefont{Sch{\"a}ffer}},
  \bibinfo{author}{\bibfnamefont{K.}~\bibnamefont{Beloy}},
  \bibinfo{author}{\bibfnamefont{D.}~\bibnamefont{Nicolodi}},
  \bibinfo{author}{\bibfnamefont{R.}~\bibnamefont{Brown}},
  \bibinfo{author}{\bibfnamefont{N.}~\bibnamefont{Hinkley}},
  \bibinfo{author}{\bibfnamefont{G.}~\bibnamefont{Milani}},
  \bibinfo{author}{\bibfnamefont{M.}~\bibnamefont{Schioppo}},
  \bibnamefont{et~al.}, \bibinfo{journal}{Nature}
  \textbf{\bibinfo{volume}{564}}, \bibinfo{pages}{87} (\bibinfo{year}{2018}).

\bibitem[{\citenamefont{Nicholson et~al.}(2015)\citenamefont{Nicholson,
  Campbell, Hutson, Marti, Bloom, McNally, Zhang, Barrett, Safronova, Strouse
  et~al.}}]{nicholson2015systematic}
\bibinfo{author}{\bibfnamefont{T.~L.} \bibnamefont{Nicholson}},
  \bibinfo{author}{\bibfnamefont{S.}~\bibnamefont{Campbell}},
  \bibinfo{author}{\bibfnamefont{R.}~\bibnamefont{Hutson}},
  \bibinfo{author}{\bibfnamefont{G.~E.} \bibnamefont{Marti}},
  \bibinfo{author}{\bibfnamefont{B.}~\bibnamefont{Bloom}},
  \bibinfo{author}{\bibfnamefont{R.~L.} \bibnamefont{McNally}},
  \bibinfo{author}{\bibfnamefont{W.}~\bibnamefont{Zhang}},
  \bibinfo{author}{\bibfnamefont{M.}~\bibnamefont{Barrett}},
  \bibinfo{author}{\bibfnamefont{M.~S.} \bibnamefont{Safronova}},
  \bibinfo{author}{\bibfnamefont{G.}~\bibnamefont{Strouse}},
  \bibnamefont{et~al.}, \bibinfo{journal}{Nature commun.}
  \textbf{\bibinfo{volume}{6}}, \bibinfo{pages}{6896} (\bibinfo{year}{2015}).

\bibitem[{\citenamefont{Nemitz et~al.}(2016)\citenamefont{Nemitz, Ohkubo,
  Takamoto, Ushijima, Das, Ohmae, and Katori}}]{nemitz2016frequency}
\bibinfo{author}{\bibfnamefont{N.}~\bibnamefont{Nemitz}},
  \bibinfo{author}{\bibfnamefont{T.}~\bibnamefont{Ohkubo}},
  \bibinfo{author}{\bibfnamefont{M.}~\bibnamefont{Takamoto}},
  \bibinfo{author}{\bibfnamefont{I.}~\bibnamefont{Ushijima}},
  \bibinfo{author}{\bibfnamefont{M.}~\bibnamefont{Das}},
  \bibinfo{author}{\bibfnamefont{N.}~\bibnamefont{Ohmae}}, \bibnamefont{and}
  \bibinfo{author}{\bibfnamefont{H.}~\bibnamefont{Katori}},
  \bibinfo{journal}{Nature Photonics} \textbf{\bibinfo{volume}{10}},
  \bibinfo{pages}{258} (\bibinfo{year}{2016}).

\bibitem[{\citenamefont{Bloom et~al.}(2014)\citenamefont{Bloom, Nicholson,
  Williams, Campbell, Bishof, Zhang, Zhang, Bromley, and
  Ye}}]{bloom2014optical}
\bibinfo{author}{\bibfnamefont{B.}~\bibnamefont{Bloom}},
  \bibinfo{author}{\bibfnamefont{T.}~\bibnamefont{Nicholson}},
  \bibinfo{author}{\bibfnamefont{J.}~\bibnamefont{Williams}},
  \bibinfo{author}{\bibfnamefont{S.}~\bibnamefont{Campbell}},
  \bibinfo{author}{\bibfnamefont{M.}~\bibnamefont{Bishof}},
  \bibinfo{author}{\bibfnamefont{X.}~\bibnamefont{Zhang}},
  \bibinfo{author}{\bibfnamefont{W.}~\bibnamefont{Zhang}},
  \bibinfo{author}{\bibfnamefont{S.}~\bibnamefont{Bromley}}, \bibnamefont{and}
  \bibinfo{author}{\bibfnamefont{J.}~\bibnamefont{Ye}},
  \bibinfo{journal}{Nature} \textbf{\bibinfo{volume}{506}}, \bibinfo{pages}{71}
  (\bibinfo{year}{2014}).

\bibitem[{\citenamefont{Middelmann et~al.}(2012)\citenamefont{Middelmann,
  Falke, Lisdat, and Sterr}}]{middelmann2012high}
\bibinfo{author}{\bibfnamefont{T.}~\bibnamefont{Middelmann}},
  \bibinfo{author}{\bibfnamefont{S.}~\bibnamefont{Falke}},
  \bibinfo{author}{\bibfnamefont{C.}~\bibnamefont{Lisdat}}, \bibnamefont{and}
  \bibinfo{author}{\bibfnamefont{U.}~\bibnamefont{Sterr}},
  \bibinfo{journal}{Phys. Rev. Lett.} \textbf{\bibinfo{volume}{109}},
  \bibinfo{pages}{263004} (\bibinfo{year}{2012}).

\bibitem[{\citenamefont{Ludlow et~al.}(2018)\citenamefont{Ludlow, McGrew,
  Zhang, Nicolodi, Fasano, Schaffer, Brown, Fox, Hinkley, Yoon
  et~al.}}]{ludlow2018optical}
\bibinfo{author}{\bibfnamefont{A.~D.} \bibnamefont{Ludlow}},
  \bibinfo{author}{\bibfnamefont{W.~F.} \bibnamefont{McGrew}},
  \bibinfo{author}{\bibfnamefont{X.}~\bibnamefont{Zhang}},
  \bibinfo{author}{\bibfnamefont{D.}~\bibnamefont{Nicolodi}},
  \bibinfo{author}{\bibfnamefont{R.~J.} \bibnamefont{Fasano}},
  \bibinfo{author}{\bibfnamefont{S.~A.} \bibnamefont{Schaffer}},
  \bibinfo{author}{\bibfnamefont{R.~C.} \bibnamefont{Brown}},
  \bibinfo{author}{\bibfnamefont{R.~W.} \bibnamefont{Fox}},
  \bibinfo{author}{\bibfnamefont{N.}~\bibnamefont{Hinkley}},
  \bibinfo{author}{\bibfnamefont{T.~H.} \bibnamefont{Yoon}},
  \bibnamefont{et~al.}, in \emph{\bibinfo{booktitle}{2018 Conference on
  Precision Electromagnetic Measurements (CPEM 2018)}}
  (\bibinfo{organization}{IEEE}, \bibinfo{year}{2018}), pp.
  \bibinfo{pages}{1--2}.

\bibitem[{\citenamefont{Lisdat et~al.}(2021)\citenamefont{Lisdat, D\"orscher,
  Nosske, and Sterr}}]{PhysRevResearch.3.L042036}
\bibinfo{author}{\bibfnamefont{C.}~\bibnamefont{Lisdat}},
  \bibinfo{author}{\bibfnamefont{S.}~\bibnamefont{D\"orscher}},
  \bibinfo{author}{\bibfnamefont{I.}~\bibnamefont{Nosske}}, \bibnamefont{and}
  \bibinfo{author}{\bibfnamefont{U.}~\bibnamefont{Sterr}},
  \bibinfo{journal}{Phys. Rev. Res.} \textbf{\bibinfo{volume}{3}},
  \bibinfo{pages}{L042036} (\bibinfo{year}{2021}).

\bibitem[{\citenamefont{Berengut et~al.}(2010)\citenamefont{Berengut, Dzuba,
  and Flambaum}}]{berengut2010enhanced}
\bibinfo{author}{\bibfnamefont{J.}~\bibnamefont{Berengut}},
  \bibinfo{author}{\bibfnamefont{V.}~\bibnamefont{Dzuba}}, \bibnamefont{and}
  \bibinfo{author}{\bibfnamefont{V.}~\bibnamefont{Flambaum}},
  \bibinfo{journal}{Phys. Rev. Lett.} \textbf{\bibinfo{volume}{105}},
  \bibinfo{pages}{120801} (\bibinfo{year}{2010}).

\bibitem[{\citenamefont{Berengut et~al.}(2011)\citenamefont{Berengut, Dzuba,
  Flambaum, and Ong}}]{berengut2011electron}
\bibinfo{author}{\bibfnamefont{J.}~\bibnamefont{Berengut}},
  \bibinfo{author}{\bibfnamefont{V.}~\bibnamefont{Dzuba}},
  \bibinfo{author}{\bibfnamefont{V.}~\bibnamefont{Flambaum}}, \bibnamefont{and}
  \bibinfo{author}{\bibfnamefont{A.}~\bibnamefont{Ong}},
  \bibinfo{journal}{Phys. Rev. Lett.} \textbf{\bibinfo{volume}{106}},
  \bibinfo{pages}{210802} (\bibinfo{year}{2011}).

\bibitem[{\citenamefont{Berengut et~al.}(2012)\citenamefont{Berengut, Dzuba,
  Flambaum, and Ong}}]{berengut2012optical}
\bibinfo{author}{\bibfnamefont{J.}~\bibnamefont{Berengut}},
  \bibinfo{author}{\bibfnamefont{V.}~\bibnamefont{Dzuba}},
  \bibinfo{author}{\bibfnamefont{V.}~\bibnamefont{Flambaum}}, \bibnamefont{and}
  \bibinfo{author}{\bibfnamefont{A.}~\bibnamefont{Ong}},
  \bibinfo{journal}{Phys. Rev. Lett.} \textbf{\bibinfo{volume}{109}},
  \bibinfo{pages}{070802} (\bibinfo{year}{2012}).

\bibitem[{\citenamefont{Kozlov et~al.}(2018)\citenamefont{Kozlov, Safronova,
  L{\'o}pez-Urrutia, and Schmidt}}]{kozlov2018highly}
\bibinfo{author}{\bibfnamefont{M.}~\bibnamefont{Kozlov}},
  \bibinfo{author}{\bibfnamefont{M.}~\bibnamefont{Safronova}},
  \bibinfo{author}{\bibfnamefont{J.~C.} \bibnamefont{L{\'o}pez-Urrutia}},
  \bibnamefont{and} \bibinfo{author}{\bibfnamefont{P.}~\bibnamefont{Schmidt}},
  \bibinfo{journal}{Rev. Mod. Phys.} \textbf{\bibinfo{volume}{90}},
  \bibinfo{pages}{045005} (\bibinfo{year}{2018}).

\bibitem[{\citenamefont{Yu et~al.}(2023)\citenamefont{Yu, Sahoo, and
  Suo}}]{Sahoo}
\bibinfo{author}{\bibfnamefont{Y.-M.} \bibnamefont{Yu}},
  \bibinfo{author}{\bibfnamefont{B.~K.} \bibnamefont{Sahoo}}, \bibnamefont{and}
  \bibinfo{author}{\bibfnamefont{B.-B.} \bibnamefont{Suo}},
  \bibinfo{journal}{Fronties in Physics} \textbf{\bibinfo{volume}{11}},
  \bibinfo{pages}{1104848} (\bibinfo{year}{2023}).

\bibitem[{\citenamefont{Dzuba et~al.}(1999{\natexlab{a}})\citenamefont{Dzuba,
  Flambaum, and Webb}}]{dzuba1999space}
\bibinfo{author}{\bibfnamefont{V.}~\bibnamefont{Dzuba}},
  \bibinfo{author}{\bibfnamefont{V.}~\bibnamefont{Flambaum}}, \bibnamefont{and}
  \bibinfo{author}{\bibfnamefont{J.}~\bibnamefont{Webb}},
  \bibinfo{journal}{Phys. Rev. Lett.} \textbf{\bibinfo{volume}{82}},
  \bibinfo{pages}{888} (\bibinfo{year}{1999}{\natexlab{a}}).

\bibitem[{\citenamefont{Dzuba et~al.}(1999{\natexlab{b}})\citenamefont{Dzuba,
  Flambaum, and Webb}}]{dzuba1999calculations}
\bibinfo{author}{\bibfnamefont{V.}~\bibnamefont{Dzuba}},
  \bibinfo{author}{\bibfnamefont{V.}~\bibnamefont{Flambaum}}, \bibnamefont{and}
  \bibinfo{author}{\bibfnamefont{J.}~\bibnamefont{Webb}},
  \bibinfo{journal}{Phys. Rev. A} \textbf{\bibinfo{volume}{59}},
  \bibinfo{pages}{230} (\bibinfo{year}{1999}{\natexlab{b}}).

\bibitem[{\citenamefont{Rosenband et~al.}(2008)\citenamefont{Rosenband, Hume,
  Schmidt, Chou, Brusch, Lorini, Oskay, Drullinger, Fortier, Stalnaker
  et~al.}}]{Science.319}
\bibinfo{author}{\bibfnamefont{T.}~\bibnamefont{Rosenband}},
  \bibinfo{author}{\bibfnamefont{D.~B.} \bibnamefont{Hume}},
  \bibinfo{author}{\bibfnamefont{P.~O.} \bibnamefont{Schmidt}},
  \bibinfo{author}{\bibfnamefont{C.~W.} \bibnamefont{Chou}},
  \bibinfo{author}{\bibfnamefont{A.}~\bibnamefont{Brusch}},
  \bibinfo{author}{\bibfnamefont{L.}~\bibnamefont{Lorini}},
  \bibinfo{author}{\bibfnamefont{W.~H.} \bibnamefont{Oskay}},
  \bibinfo{author}{\bibfnamefont{R.~E.} \bibnamefont{Drullinger}},
  \bibinfo{author}{\bibfnamefont{T.~M.} \bibnamefont{Fortier}},
  \bibinfo{author}{\bibfnamefont{J.~E.} \bibnamefont{Stalnaker}},
  \bibnamefont{et~al.}, \bibinfo{journal}{Science}
  \textbf{\bibinfo{volume}{319}}, \bibinfo{pages}{1808} (\bibinfo{year}{2008}).

\bibitem[{\citenamefont{Berengut et~al.}(2018)\citenamefont{Berengut, Budker,
  Delaunay, Flambaum, Frugiuele, Fuchs, Grojean, Harnik, Ozeri, Perez
  et~al.}}]{berengut2018probing}
\bibinfo{author}{\bibfnamefont{J.~C.} \bibnamefont{Berengut}},
  \bibinfo{author}{\bibfnamefont{D.}~\bibnamefont{Budker}},
  \bibinfo{author}{\bibfnamefont{C.}~\bibnamefont{Delaunay}},
  \bibinfo{author}{\bibfnamefont{V.~V.} \bibnamefont{Flambaum}},
  \bibinfo{author}{\bibfnamefont{C.}~\bibnamefont{Frugiuele}},
  \bibinfo{author}{\bibfnamefont{E.}~\bibnamefont{Fuchs}},
  \bibinfo{author}{\bibfnamefont{C.}~\bibnamefont{Grojean}},
  \bibinfo{author}{\bibfnamefont{R.}~\bibnamefont{Harnik}},
  \bibinfo{author}{\bibfnamefont{R.}~\bibnamefont{Ozeri}},
  \bibinfo{author}{\bibfnamefont{G.}~\bibnamefont{Perez}},
  \bibnamefont{et~al.}, \bibinfo{journal}{Phys. Rev. Lett.}
  \textbf{\bibinfo{volume}{120}}, \bibinfo{pages}{091801}
  (\bibinfo{year}{2018}).

\bibitem[{\citenamefont{Flambaum et~al.}(2018)\citenamefont{Flambaum, Geddes,
  and Viatkina}}]{flambaum2018isotope}
\bibinfo{author}{\bibfnamefont{V.}~\bibnamefont{Flambaum}},
  \bibinfo{author}{\bibfnamefont{A.}~\bibnamefont{Geddes}}, \bibnamefont{and}
  \bibinfo{author}{\bibfnamefont{A.}~\bibnamefont{Viatkina}},
  \bibinfo{journal}{Phys. Rev. A} \textbf{\bibinfo{volume}{97}},
  \bibinfo{pages}{032510} (\bibinfo{year}{2018}).

\bibitem[{\citenamefont{Delaunay et~al.}(2017)\citenamefont{Delaunay, Ozeri,
  Perez, and Soreq}}]{delaunay2017probing}
\bibinfo{author}{\bibfnamefont{C.}~\bibnamefont{Delaunay}},
  \bibinfo{author}{\bibfnamefont{R.}~\bibnamefont{Ozeri}},
  \bibinfo{author}{\bibfnamefont{G.}~\bibnamefont{Perez}}, \bibnamefont{and}
  \bibinfo{author}{\bibfnamefont{Y.}~\bibnamefont{Soreq}},
  \bibinfo{journal}{Phys. Rev. D} \textbf{\bibinfo{volume}{96}},
  \bibinfo{pages}{093001} (\bibinfo{year}{2017}).

\bibitem[{\citenamefont{Yudin et~al.}(2011)\citenamefont{Yudin, Taichenachev,
  Okhapkin, Bagayev, Tamm, Peik, Huntemann, Mehlst{\"a}ubler, and
  Riehle}}]{yudin2011atomic}
\bibinfo{author}{\bibfnamefont{V.}~\bibnamefont{Yudin}},
  \bibinfo{author}{\bibfnamefont{A.}~\bibnamefont{Taichenachev}},
  \bibinfo{author}{\bibfnamefont{M.}~\bibnamefont{Okhapkin}},
  \bibinfo{author}{\bibfnamefont{S.}~\bibnamefont{Bagayev}},
  \bibinfo{author}{\bibfnamefont{C.}~\bibnamefont{Tamm}},
  \bibinfo{author}{\bibfnamefont{E.}~\bibnamefont{Peik}},
  \bibinfo{author}{\bibfnamefont{N.}~\bibnamefont{Huntemann}},
  \bibinfo{author}{\bibfnamefont{T.}~\bibnamefont{Mehlst{\"a}ubler}},
  \bibnamefont{and} \bibinfo{author}{\bibfnamefont{F.}~\bibnamefont{Riehle}},
  \bibinfo{journal}{Phys. Rev. Lett.} \textbf{\bibinfo{volume}{107}},
  \bibinfo{pages}{030801} (\bibinfo{year}{2011}).

\bibitem[{\citenamefont{Dzuba}(2014)}]{dzuba2014combination}
\bibinfo{author}{\bibfnamefont{V.}~\bibnamefont{Dzuba}},
  \bibinfo{journal}{Phys. Rev. A} \textbf{\bibinfo{volume}{90}},
  \bibinfo{pages}{012517} (\bibinfo{year}{2014}).

\bibitem[{\citenamefont{Dzuba et~al.}(2017)\citenamefont{Dzuba, Berengut,
  Harabati, and Flambaum}}]{dzuba2017combining}
\bibinfo{author}{\bibfnamefont{V.}~\bibnamefont{Dzuba}},
  \bibinfo{author}{\bibfnamefont{J.}~\bibnamefont{Berengut}},
  \bibinfo{author}{\bibfnamefont{C.}~\bibnamefont{Harabati}}, \bibnamefont{and}
  \bibinfo{author}{\bibfnamefont{V.}~\bibnamefont{Flambaum}},
  \bibinfo{journal}{Phys. Rev. A} \textbf{\bibinfo{volume}{95}},
  \bibinfo{pages}{012503} (\bibinfo{year}{2017}).

\bibitem[{\citenamefont{Dzuba}(2005)}]{dzuba2005v}
\bibinfo{author}{\bibfnamefont{V.}~\bibnamefont{Dzuba}},
  \bibinfo{journal}{Phys. Rev. A} \textbf{\bibinfo{volume}{71}},
  \bibinfo{pages}{032512} (\bibinfo{year}{2005}).

\bibitem[{\citenamefont{Dzuba et~al.}(2006)\citenamefont{Dzuba, Flambaum, and
  Safronova}}]{DzuFlaSaf06}
\bibinfo{author}{\bibfnamefont{V.~A.} \bibnamefont{Dzuba}},
  \bibinfo{author}{\bibfnamefont{V.~V.} \bibnamefont{Flambaum}},
  \bibnamefont{and} \bibinfo{author}{\bibfnamefont{M.~S.}
  \bibnamefont{Safronova}}, \bibinfo{journal}{Phys. Rev. A}
  \textbf{\bibinfo{volume}{73}}, \bibinfo{pages}{022112}
  (\bibinfo{year}{2006}).

\bibitem[{\citenamefont{Flambaum and Ginges}(2005)}]{radpot}
\bibinfo{author}{\bibfnamefont{V.~V.} \bibnamefont{Flambaum}} \bibnamefont{and}
  \bibinfo{author}{\bibfnamefont{J.~S.~M.} \bibnamefont{Ginges}},
  \bibinfo{journal}{Phys. Rev. A} \textbf{\bibinfo{volume}{72}},
  \bibinfo{pages}{052115} (\bibinfo{year}{2005}).

\bibitem[{\citenamefont{Johnson and Sapirstein}(1986)}]{johnson1986computation}
\bibinfo{author}{\bibfnamefont{W.}~\bibnamefont{Johnson}} \bibnamefont{and}
  \bibinfo{author}{\bibfnamefont{J.}~\bibnamefont{Sapirstein}},
  \bibinfo{journal}{Phys. Rev. Lett.} \textbf{\bibinfo{volume}{57}},
  \bibinfo{pages}{1126} (\bibinfo{year}{1986}).

\bibitem[{\citenamefont{Johnson et~al.}(1988)\citenamefont{Johnson, Blundell,
  and Sapirstein}}]{johnson1988finite}
\bibinfo{author}{\bibfnamefont{W.~R.} \bibnamefont{Johnson}},
  \bibinfo{author}{\bibfnamefont{S.~A.} \bibnamefont{Blundell}},
  \bibnamefont{and}
  \bibinfo{author}{\bibfnamefont{J.}~\bibnamefont{Sapirstein}},
  \bibinfo{journal}{Phys. Rev. A} \textbf{\bibinfo{volume}{37}},
  \bibinfo{pages}{307} (\bibinfo{year}{1988}).

\bibitem[{\citenamefont{Dzuba et~al.}(1987)\citenamefont{Dzuba, Flambaum,
  Silvestrov, and Sushkov}}]{dzuba1987correlation}
\bibinfo{author}{\bibfnamefont{V.}~\bibnamefont{Dzuba}},
  \bibinfo{author}{\bibfnamefont{V.}~\bibnamefont{Flambaum}},
  \bibinfo{author}{\bibfnamefont{P.}~\bibnamefont{Silvestrov}},
  \bibnamefont{and} \bibinfo{author}{\bibfnamefont{O.}~\bibnamefont{Sushkov}},
  \bibinfo{journal}{J. Phys. B} \textbf{\bibinfo{volume}{20}},
  \bibinfo{pages}{1399} (\bibinfo{year}{1987}).

\bibitem[{\citenamefont{Dzuba et~al.}(1996)\citenamefont{Dzuba, Flambaum, and
  Kozlov}}]{dzuba1996combination}
\bibinfo{author}{\bibfnamefont{V.~A.} \bibnamefont{Dzuba}},
  \bibinfo{author}{\bibfnamefont{V.~V.} \bibnamefont{Flambaum}},
  \bibnamefont{and} \bibinfo{author}{\bibfnamefont{M.~G.}
  \bibnamefont{Kozlov}}, \bibinfo{journal}{Phys. Rev. A}
  \textbf{\bibinfo{volume}{54}}, \bibinfo{pages}{3948} (\bibinfo{year}{1996}).

\bibitem[{\citenamefont{Allehabi
  et~al.}(2022{\natexlab{a}})\citenamefont{Allehabi, Dzuba, and
  Flambaum}}]{allehabi2022atomic}
\bibinfo{author}{\bibfnamefont{S.~O.} \bibnamefont{Allehabi}},
  \bibinfo{author}{\bibfnamefont{V.}~\bibnamefont{Dzuba}}, \bibnamefont{and}
  \bibinfo{author}{\bibfnamefont{V.}~\bibnamefont{Flambaum}},
  \bibinfo{journal}{Phys. Rev. A} \textbf{\bibinfo{volume}{106}},
  \bibinfo{pages}{032807} (\bibinfo{year}{2022}{\natexlab{a}}).

\bibitem[{\citenamefont{Allehabi
  et~al.}(2022{\natexlab{b}})\citenamefont{Allehabi, Brewer, Dzuba, Flambaum,
  and Beloy}}]{allehabi2022high}
\bibinfo{author}{\bibfnamefont{S.~O.} \bibnamefont{Allehabi}},
  \bibinfo{author}{\bibfnamefont{S.}~\bibnamefont{Brewer}},
  \bibinfo{author}{\bibfnamefont{V.}~\bibnamefont{Dzuba}},
  \bibinfo{author}{\bibfnamefont{V.}~\bibnamefont{Flambaum}}, \bibnamefont{and}
  \bibinfo{author}{\bibfnamefont{K.}~\bibnamefont{Beloy}},
  \bibinfo{journal}{Phys. Rev. A} \textbf{\bibinfo{volume}{106}},
  \bibinfo{pages}{043101} (\bibinfo{year}{2022}{\natexlab{b}}).

\bibitem[{\citenamefont{Allehabi et~al.}(2023)\citenamefont{Allehabi, Dzuba,
  and Flambaum}}]{allehabi2023prospects}
\bibinfo{author}{\bibfnamefont{S.~O.} \bibnamefont{Allehabi}},
  \bibinfo{author}{\bibfnamefont{V.}~\bibnamefont{Dzuba}}, \bibnamefont{and}
  \bibinfo{author}{\bibfnamefont{V.}~\bibnamefont{Flambaum}},
  \bibinfo{journal}{Phys. Rev. A} \textbf{\bibinfo{volume}{107}},
  \bibinfo{pages}{063103} (\bibinfo{year}{2023}).

\bibitem[{\citenamefont{Schmidt et~al.}(2005)\citenamefont{Schmidt, Rosenband,
  Langer, Itano, Bergquist, and Wineland}}]{schmidt2005spectroscopy}
\bibinfo{author}{\bibfnamefont{P.~O.} \bibnamefont{Schmidt}},
  \bibinfo{author}{\bibfnamefont{T.}~\bibnamefont{Rosenband}},
  \bibinfo{author}{\bibfnamefont{C.}~\bibnamefont{Langer}},
  \bibinfo{author}{\bibfnamefont{W.~M.} \bibnamefont{Itano}},
  \bibinfo{author}{\bibfnamefont{J.~C.} \bibnamefont{Bergquist}},
  \bibnamefont{and} \bibinfo{author}{\bibfnamefont{D.~J.}
  \bibnamefont{Wineland}}, \bibinfo{journal}{Science}
  \textbf{\bibinfo{volume}{309}}, \bibinfo{pages}{749} (\bibinfo{year}{2005}).

\bibitem[{\citenamefont{Ludlow et~al.}(2015)\citenamefont{Ludlow, Boyd, Ye,
  Peik, and Schmidt}}]{ludlow2015optical}
\bibinfo{author}{\bibfnamefont{A.~D.} \bibnamefont{Ludlow}},
  \bibinfo{author}{\bibfnamefont{M.~M.} \bibnamefont{Boyd}},
  \bibinfo{author}{\bibfnamefont{J.}~\bibnamefont{Ye}},
  \bibinfo{author}{\bibfnamefont{E.}~\bibnamefont{Peik}}, \bibnamefont{and}
  \bibinfo{author}{\bibfnamefont{P.~O.} \bibnamefont{Schmidt}},
  \bibinfo{journal}{Rev. Mod. Phys.} \textbf{\bibinfo{volume}{87}},
  \bibinfo{pages}{637} (\bibinfo{year}{2015}).

\bibitem[{\citenamefont{W{\"u}bbena et~al.}(2012)\citenamefont{W{\"u}bbena,
  Amairi, Mandel, and Schmidt}}]{wubbena2012sympathetic}
\bibinfo{author}{\bibfnamefont{J.~B.} \bibnamefont{W{\"u}bbena}},
  \bibinfo{author}{\bibfnamefont{S.}~\bibnamefont{Amairi}},
  \bibinfo{author}{\bibfnamefont{O.}~\bibnamefont{Mandel}}, \bibnamefont{and}
  \bibinfo{author}{\bibfnamefont{P.~O.} \bibnamefont{Schmidt}},
  \bibinfo{journal}{Phys. Rev. A} \textbf{\bibinfo{volume}{85}},
  \bibinfo{pages}{043412} (\bibinfo{year}{2012}).

\bibitem[{\citenamefont{Dzuba}(2020)}]{dzuba2020calculation}
\bibinfo{author}{\bibfnamefont{V.}~\bibnamefont{Dzuba}},
  \bibinfo{journal}{Symmetry} \textbf{\bibinfo{volume}{12}},
  \bibinfo{pages}{1950} (\bibinfo{year}{2020}).

\bibitem[{\citenamefont{Dalgarno and Lewis}(1955)}]{dalgarno1955exact}
\bibinfo{author}{\bibfnamefont{A.}~\bibnamefont{Dalgarno}} \bibnamefont{and}
  \bibinfo{author}{\bibfnamefont{J.~T.} \bibnamefont{Lewis}},
  \bibinfo{journal}{Proc. R. Soc. A} \textbf{\bibinfo{volume}{233}},
  \bibinfo{pages}{70} (\bibinfo{year}{1955}).

\bibitem[{\citenamefont{Itano}(2000)}]{itano2000external}
\bibinfo{author}{\bibfnamefont{W.~M.} \bibnamefont{Itano}},
  \bibinfo{journal}{J. Res. Natl. Inst. Stand. Technol.}
  \textbf{\bibinfo{volume}{105}}, \bibinfo{pages}{829} (\bibinfo{year}{2000}).

\bibitem[{\citenamefont{Flambaum and Dzuba}(2009)}]{FlaDzuCJP}
\bibinfo{author}{\bibfnamefont{V.~V.} \bibnamefont{Flambaum}} \bibnamefont{and}
  \bibinfo{author}{\bibfnamefont{V.~A.} \bibnamefont{Dzuba}},
  \bibinfo{journal}{Can. J. Phys.} \textbf{\bibinfo{volume}{87}},
  \bibinfo{pages}{25} (\bibinfo{year}{2009}).

\end{thebibliography}

\end{document}